%% file: GPR_arxiv26.tex
\documentclass[sigconf]{acmart}
\usepackage{multirow}
\usepackage{float}  
\usepackage{graphicx}
\usepackage{subcaption}
\usepackage{caption}
\usepackage{booktabs}
\usepackage{siunitx}
\usepackage{threeparttable}
\usepackage{adjustbox}
\usepackage{multirow}
\usepackage{xcolor}
\usepackage[absolute,overlay]{textpos}
\newcommand{\bb}{\hspace{0mm} $\bullet$ \hspace{0.8mm}}

\AtBeginDocument{%
  }

\usepackage[ruled,vlined]{algorithm2e}
\usepackage{amsmath}
\setcopyright{acmlicensed}
\copyrightyear{2018}
\acmYear{2018}
\acmDOI{XXXXXXX.XXXXXXX}
\acmISBN{978-1-4503-XXXX-X/2018/06}

\begin{document}
\begin{textblock*}{\paperwidth}(1.5cm, 1.2cm)
\includegraphics[height=0.7cm]{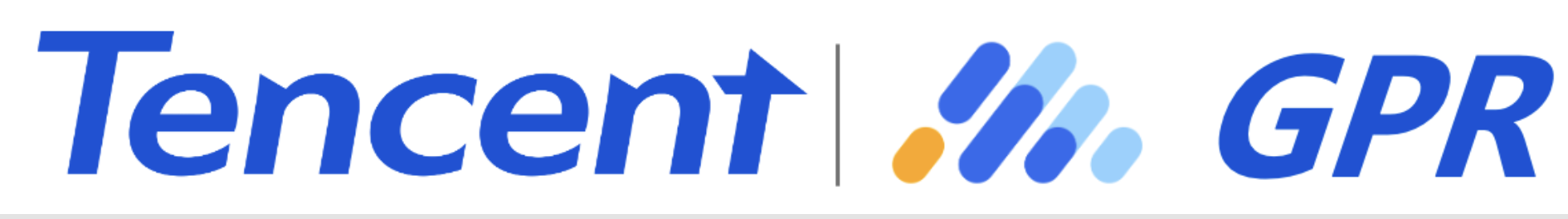} 
\end{textblock*}

\title{GPR: Towards a Generative Pre-trained One-Model Paradigm for Large-Scale Advertising Recommendation}

\author{{\Large Jun Zhang $^{1*+}$, Yi Li $^{1*}$, Yue Liu $^{1*}$, Changping Wang $^{1*}$, Yuan Wang $^{1*}$, Yuling Xiong $^{1*}$, Xun Liu $^{1*}$, Haiyang Wu $^{1*}$, Qian Li $^{1*}$, Enming Zhang $^1$, Jiawei Sun $^1$, Xin Xu $^1$, Zishuai Zhang $^1$, Ruoran Liu $^1$, Suyuan Huang $^1$, Zhaoxin Zhang $^1$, Zhengkai Guo $^1$, Shuojin Yang $^2$, Meng-Hao Guo $^2$, Huan Yu $^1$, Jie Jiang $^1$, Shi-Min Hu $^2$}}

\affiliation{
  \institution{$^1$ Tencent Inc., China; $^2$ Tsinghua University, China}
  \country{}
}


\renewcommand{\shortauthors}{Zhang et al.}

\begin{abstract}

As an intelligent infrastructure connecting users with commercial content, advertising recommendation systems play a central role in information flow and value creation within the digital economy. However, existing multi-stage advertising recommendation systems suffer from objective misalignment and error propagation, making it difficult to achieve global optimality, while unified generative recommendation models still struggle to meet the demands of practical industrial applications.
To address these issues, we propose GPR (Generative Pre-trained Recommender) — the first one-model framework that redefines advertising recommendation as an end-to-end generative task, replacing the traditional cascading paradigm with a unified generative approach. To realize GPR, we introduce three key innovations spanning unified representation, network architecture, and training strategy.
First, we design a unified input schema and tokenization method tailored to advertising scenarios, mapping both ads and organic content into a shared multi-level semantic ID space, thereby enhancing semantic alignment and modeling consistency across heterogeneous data.
Second, we develop the Heterogeneous Hierarchical Decoder (HHD), a dual-decoder architecture that decouples user intent modeling from ad generation, achieving a balance between training efficiency and inference flexibility while maintaining strong modeling capacity.
Finally, we propose a multi-stage joint training strategy that integrates Multi-Token Prediction (MTP), Value-Aware Fine-Tuning and the Hierarchy Enhanced Policy Optimization (HEPO) algorithm, forming a complete generative recommendation pipeline that unifies interest modeling, value alignment, and policy optimization. GPR has been fully deployed in the Tencent Weixin Channels advertising system, delivering significant improvements in key business metrics including Gross Merchandise Volume (GMV) and CTCVR, clearly demonstrating that GPR maintains strong competitiveness against a highly optimized and mature cascading system.

\end{abstract}

\ccsdesc[500]{Information systems~Recommender systems}

\keywords{Recommender Systems, Generative Modeling, Advertising}



\maketitle

\def\thefootnote{*}\footnotetext{These authors contributed equally to this work}
\def\thefootnote{+}\footnotetext{Corresponding author}

\begin{figure}
    \centering
    \includegraphics[width=1.0\linewidth]{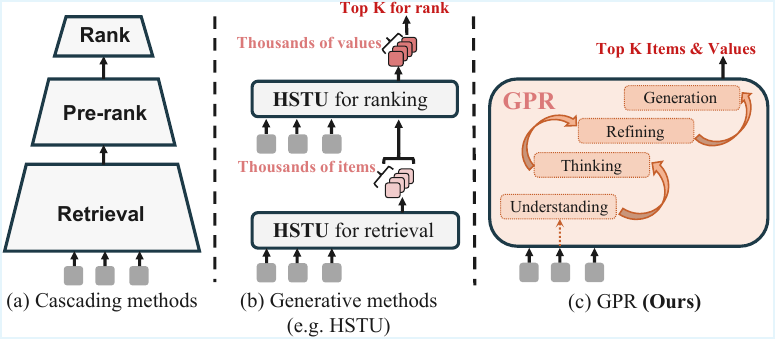}
    \caption{Comparison between previous methods and ours.}
    \label{fig:teaser}
\end{figure}

\input{1_Introduction_section}
\input{3_GPR_v2}
\input{4_Training_GPR_section_v2}

\input{5_Experiments_section}

\input{2_Related_Work_section_v2}
\input{6_Conclusion_section}

\begin{acks}
We'd like to express our sincere gratitude to the following individuals (alphabetical order) for their invaluable contributions: Chenchen Yu, Dekai Sun, He Wei, Jiacheng Han, Jiafan Zhou, Jingwen Wang, Junbang Huo, Juntong Yan, Ke Cheng, Longfei Lu, Roger Jin, Siao Li, Wei Cao, Yankuan Liang, Yifei Liu, Yihang Su, Yimin Wang, Zhenmao Li, Zhennan Pang.
\end{acks}

\bibliographystyle{ACM-Reference-Format}
\bibliography{sample-base}



\end{document}

%% file: 1_Introduction_section.tex
\section{Introduction}

Online advertising recommendation is a vital component of the digital economy, 
with its core task being to precisely match the right ads to the right users based on their historical behaviors. 
Such systems must operate under strict real-time, low-latency requirements, serving hundreds of millions of users and tens of millions of dynamic ads. Consequently, the timeliness and stability of system performance directly determine the functioning of a multi-billion-dollar ecosystem. Achieving a dynamic balance among user experience, advertiser return on investment (ROI), and platform revenue has become a key challenge in this field.

\begin{figure*}
    \centering
    \includegraphics[width=1.\linewidth]{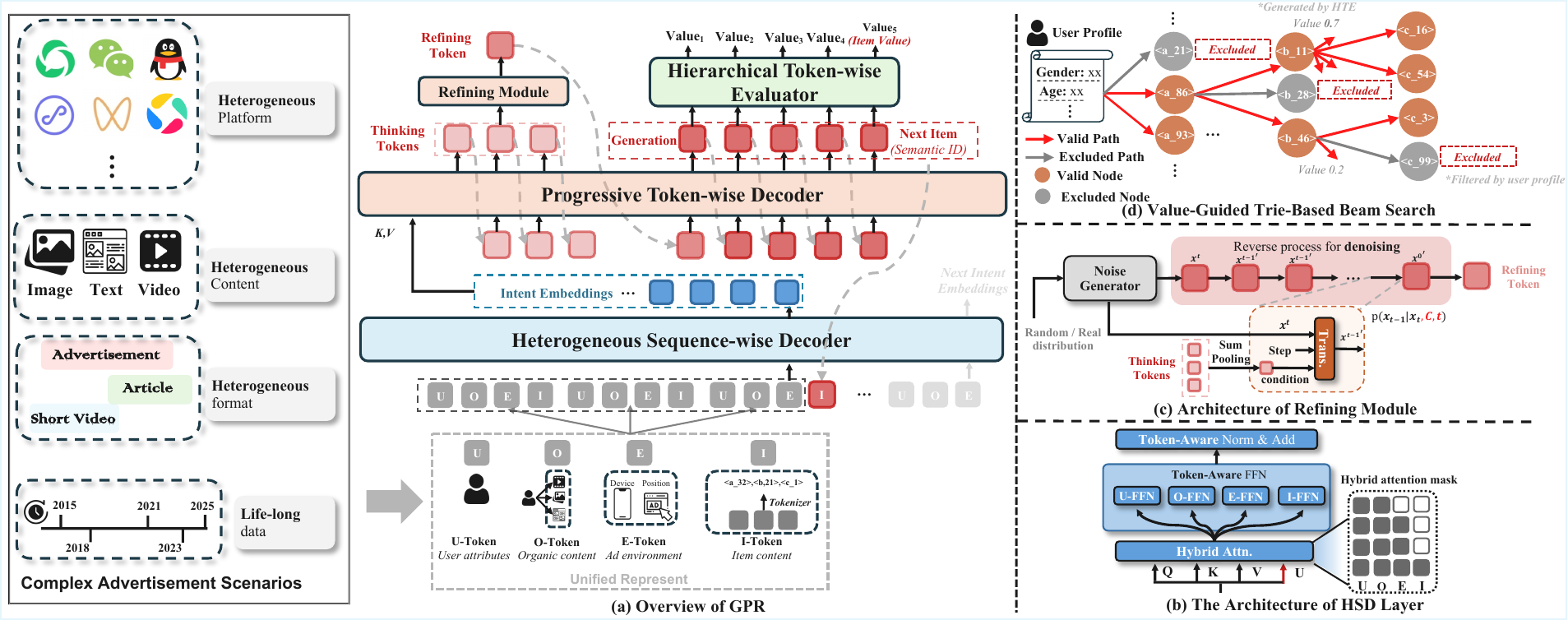}
    \caption{Overall Architecture of GPR.}
    \label{fig:gpr_v3_1}
\end{figure*}

Traditional advertising recommendation systems typically adopt a multi-stage cascading pipeline of “retrieval–pre-ranking–ranking”

\noindent~\cite{su2009survey,huang2013learning,cheng2016wide,zhou2018deep,zhang2024wukong,zhu2025rankmixer}, as shown in Fig.~\ref{fig:teaser}(a). However, such pipelines suffer from inconsistent optimization objectives across stages: the retrieval stage emphasizes coverage optimization, while the final ranking stage focuses on business outcome prediction. This misalignment prevents the system from achieving global optimality. Moreover, the limited representational capacity of early retrieval models often leads to the premature elimination of potentially high-quality candidates, restricting subsequent ranking models from fully exploiting deep feature interactions and resulting in an information bottleneck. In addition, the cascading architecture relies on highly complex engineering implementations, requiring substantial resources to maintain cross-stage consistency, which in turn hinders rapid algorithmic iteration and system scalability.

In recent years, with the rapid advancement of large language models (LLMs) ~\cite{vaswani2017attention,achiam2023gpt,guo2025deepseek}, particularly their breakthroughs in generative capabilities~\cite{zhai2024actions,rajput2023recommender}, the research paradigm of recommendation systems has been continuously evolving. HSTU~\cite{zhai2024actions} introduced generative modeling into the traditional cascading architecture, as shown in Fig.~\ref{fig:teaser}(b), further driving the transition from the multi-stage pipeline to a unified generative recommendation framework~\cite{zhou2025onerec,han2025mtgr}. Such models can directly generate optimal recommendations based on a holistic understanding of user interests and semantic context, thereby ensuring the consistency of optimization objectives.
However, due to the inherent complexity of advertising recommendation tasks, existing generative models still face multiple challenges in real-world industrial applications:

\textbf{1) Extreme Heterogeneity in Data and Behavior:} Ads are typically interleaved with organic content such as short videos, social feeds, and news articles, resulting in heterogeneity at both the sequence and item levels, as well as diverse user behaviors. The system must simultaneously process interactions with both ads and non-ad content (e.g., clicks, conversions, views, and reads), leading to highly noisy and complex data distributions. This extreme heterogeneity greatly increases the difficulty of user interest modeling and semantic alignment, placing stringent demands on the model’s unified representation and multimodal modeling capabilities.

\textbf{2) Efficiency–Flexibility Trade-off:} In industrial-scale advertising recommendation, the model must not only support efficient training for large-scale data updates but also provide flexible decoding capabilities to process ultra-long user behavior sequences in real time and match ads under multiple constraints (e.g., targeting, bidding, budget). Decoder-only architectures~\cite{zhai2024actions} enable efficient user-level training but suffer from limited decoding flexibility, whereas encoder–decoder architectures~\cite{zhou2025onerec} allow flexible inference but incur prohibitively high training costs due to pointwise loss objectives, making them unsuitable for real-time advertising systems.
Balancing efficient training with flexible inference thus remains a key challenge.

\textbf{3) Revenue and Multi-stakeholder Value Optimization:} Advertising recommendation systems must maximize the overall ecosystem value among multiple stakeholders — balancing user experience, advertiser return on investment (ROI), and platform revenue. Existing pre-training methods primarily rely on optimizing simplified, single objectives (e.g., maximizing predicted Click-Through Rate or Conversion Rate) in isolation. Such siloed optimization leads to inherent objective misalignment and local optimality, thus failing to achieve the desired globally optimized business value.

To address the aforementioned challenges, we propose \textbf{GPR (Generative Pre-trained Recommender)}. Through systematic innovations in unified tokenization, decoder architecture and training strategy, GPR effectively unifies heterogeneous data, achieving a balance between system efficiency and decoding flexibility, while precisely capturing business value and optimizing global multi-stakeholder objectives.
To the best of our knowledge, GPR is the first end-to-end generative advertising recommendation framework successfully deployed in a large-scale real-world advertising system, as shown in Fig.~\ref{fig:teaser}(c).
The main contributions of this work are summarized as follows:

\bb We propose a unified input schema and tokenization method, which represent the entire user journey through four types of tokens, and introduce a novel RQ-Kmeans+ quantization model to map both content and ads into a shared semantic space, enabling efficient modeling of heterogeneous and ultra-long sequential data.

\bb We propose a dual-decoder–based generative architecture, called Heterogeneous Hierarchical Decoder (HHD), which hierarchically models user understanding and recommended items generation for finer interest representation and more accurate recommendations. During decoding, HHD integrates trie constraints, value guidance, and efficient multi-stage pruning, greatly improving generation accuracy and reliability.

\bb We propose a multi-stage joint training strategy that integrates Multi-Token Prediction (MTP), Value-Aware Fine-Tuning (VAFT) and the Hierarchy Enhanced Policy Optimization (HEPO) algorithm, thereby constructing a comprehensive generative recommendation training pipeline that seamlessly unifies interest modeling, value alignment, and policy optimization.

\bb We have fully deployed GPR in the Tencent Weixin Channels advertising system. Results from large-scale online A/B testing show that it delivers a significant improvements in key business metrics including Gross Merchandise Volume (GMV) and CTCVR, clearly demonstrating that GPR maintains strong competitiveness against a highly optimized and mature cascading system.

%% file: 3_GPR_v2.tex
\section{Generative Pre-trained Recommender}

This section introduces GPR — a unified framework capable of generating both recommended items and auction value within a single generative model.
To address the noisy, highly heterogeneous, and ultra-long sequential nature of user behavior data, we first propose a novel Input Schema that represents users’ long-term behaviors as a unified and continuous sequence of tokens.
Building upon this representation, we design a new quantizer, RQ-Kmeans+, which significantly improves codebook utilization efficiency and effectively resolves the common issue of codebook collapse, while maintaining flexibility in the latent space.
Then we construct a dual-decoder-based generative framework, termed the Heterogeneous Hierarchical Decoder (HHD), which enables deep understanding of complex user behavior patterns and accurate prediction of both target items and their corresponding business values. Finally, we design a Trie-based beam search algorithm with value guidance to improve inference efficiency and performance. The overall architecture is illustrated in Fig. \ref{fig:gpr_v3_1}.

\subsection{Input Schema and Processing}
\label{sec:input}
GPR model is required to process noisy, heterogeneous and ultra-long user information, as the real-world advertising platforms typically encompass a variety of diverse scenarios, such as Weixin Channels, Moments, and Official Accounts. In addition, users’ behavioral sequences in these platforms can span over extensive periods and accumulate to an exceptionally long history, further increasing the complexity of information the model must process and understand.
Therefore, we propose a unified input schema for GPR to represent the whole journey of users, which consists of User Token (U-Token), Organic Token (O-Token), Environment Token (E-Token) and Item Token (I-Token). 
Specifically, the U-Token represents users' attributes and preferences. The O-Token encapsulates users' organic content, such as short videos and articles. Additionally, the E-Token encodes the immediate context surrounding an advertisement request, while the I-Token represents an ad item where the user has interacted.

Furthermore, as demonstrated by previous GR models~\cite{rajput2023recommender,zhou2025onerec}, quantizing item embeddings into semantic IDs can extract crucial information in items, while also generate discrete representation of items, better aligning with generative model paradigms. Therefore, we convert all contents in O-Token and items in I-Token into discrete semantic IDs. RQ-VAE~\cite{rq-vae} and RQ-Kmeans~\cite{rq-kmeans} are two popular methods. However, both of them suffer from issues such as "codebook collapse" and "insufficient robustness of the latent space", leading to low utilization of the semantic space and limited representational capacity. To address these issues, we propose a new quantization model named \textbf{RQ-Kmeans+}, which consists of an encoder, residual codebooks and a decoder.

\begin{figure}[t!]
    \centering
    \includegraphics[width=1.0\linewidth]{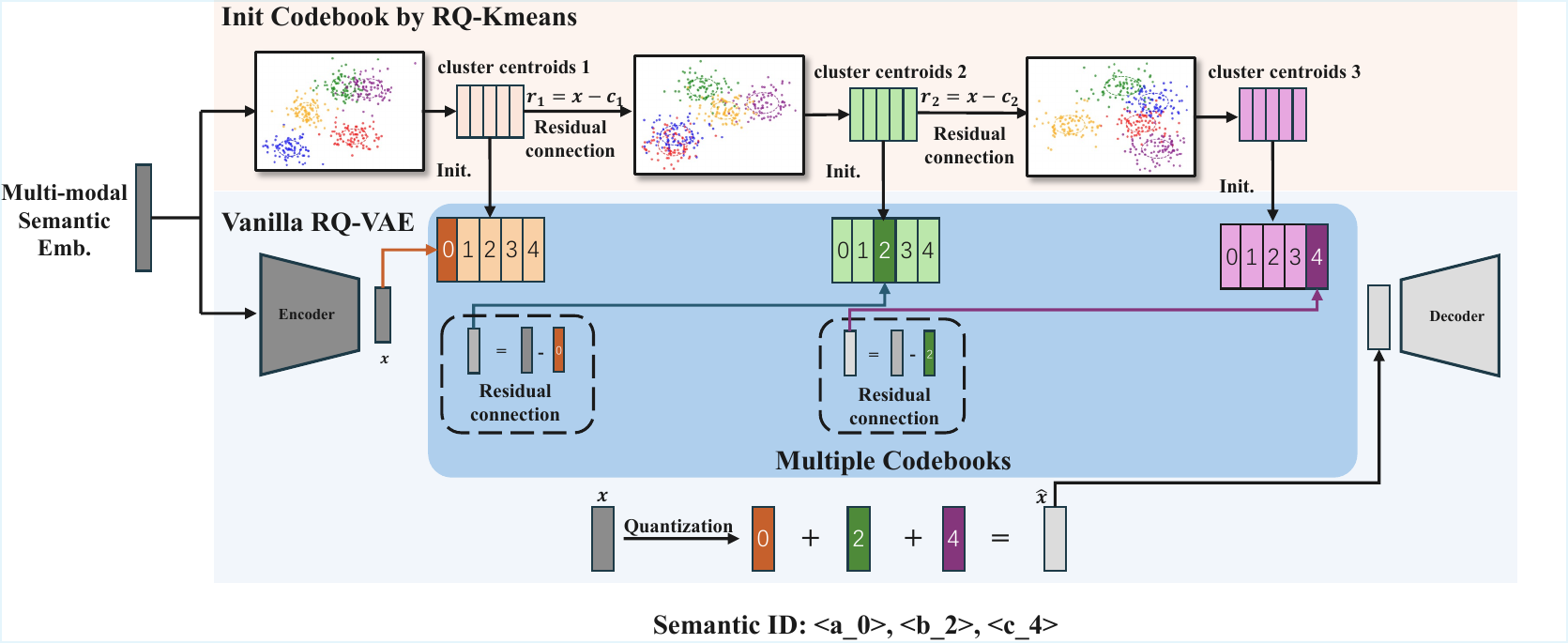}
    \caption{Overall Architecture of RQ-Kmeans+.}
    \label{fig:rq-kmeans}
\end{figure}

We attribute codebook collapse primarily to random initialization, where some vectors are rarely activated during training and eventually become “dead vectors”, resulting in inefficient codebook utilization. To address this, as shown in Fig. ~\ref{fig:rq-kmeans}, RQ-Kmeans+ first employs RQ-Kmeans to generate a high-quality codebook, which is then used as initialization weights. The codebook is subsequently updated using the same loss function as RQ-VAE, enabling adaptation to the current learnable latent space. Furthermore, we introduce a residual connection on the encoder side to ensure that the output distribution remains close to the input distribution in the early stages of training, thereby accelerating convergence and stabilizing latent-space alignment. Ultimately, RQ-Kmeans+ significantly improves codebook utilization while maintaining flexibility in the latent space, effectively resolving the collapse problem. 

\subsection{Heterogeneous Hierarchical Decoder}

We propose a decoder-only generative architecture, named \textbf{Heterogeneous Hierarchical Decoder (HHD)}, which consists of a Heterogeneous Sequence-wise Decoder (HSD) Module, a Progressive Token-wise Decoder (PTD) Module and a Hierarchical Token-wise Evaluator (HTE) Module, as shown in Fig. ~\ref{fig:gpr_v3_1} (a). This hierarchical structure  decouples the understanding of user behaviors from next item prediction, thereby enabling a finer-grained understanding of user preferences and consequently more accurate next-item prediction. 
Specifically, the HSD Module utilizes a Hybrid Attention mechanism, alongside Token-Aware Normalization and Feed-Forward Networks (FFN), to thoroughly grasp user behaviors and generate high-quality intent embeddings. Additionally, Mixture-of-Recursions (MoR) strategy is employed to reduce memory consumption and enhance the model's reasoning depth. Guided by the derived intent embeddings, PTD Module predicts next item following a \textit{"Thinking-Refining-Generation"} paradigm to facilitate more accurate prediction results by thinking tokens and refining token. 
During the generation process, the HTE module outputs estimated values for each semantic code and the final item, facilitating pruning and auction.

\paragraph{\textbf{Heterogeneous Sequence-wise Decoder (HSD)}}
\label{sec:hsd}

The HSD module is a primary decoder, which stacks HSTU blocks~\cite{zhai2024actions} and takes the unified token sequence as input to understand users' actions and generate intent embeddings. Building upon the foundational HSTU block, we introduce several critical enhancements to form the block utilized in the HSD, including the Hybrid Attention mechanism, Token-Aware Normalization, and an improved Feed Forward Network (FFN), as shown in Fig. ~\ref{fig:gpr_v3_1} (b).

Unlike standard attention mechanisms, the HSD Attention mechanism incorporates an extra embedding, $U$, which adaptively modulate the attention weights, $W$. This modulation enables the HSD to focus more effectively on relevant user behaviors while actively attenuating less informative interactions~\cite{zhai2024actions}. 
Moreover, given that the U-Token, O-Token, and E-Token in our input schema collectively function as a prompt for item prediction, applying a vanilla causal mask within this prompt region becomes unnecessarily restrictive. To address this limitation, we propose the Hybrid Attention mask, $M^{hybrid}$. Within the prefix block, tokens visible with each other freely using bi-directional attention. This design enables the model to fully exploit the contextual interplay among the prompt tokens, thereby constructing a more comprehensive context prior to making a prediction. The Hybrid Attention can be expressed as:

\vspace{-10pt}
\begin{equation}
\begin{aligned}
&\mathtt{HybridAttn}(\cdot) = Softmax\left( \frac{QK^\top}{\sqrt{d}}+ M^{\text{hybrid}} \right) V \odot U \\[1.5ex]
\end{aligned}
\end{equation}
\vspace{-10pt}
\begin{equation}
\begin{aligned}
&M_{ij}^{\text{hybrid}} =
\begin{cases}
0, & \text{if } i < j \text{ or } X_i, X_j \in \{\text{U/O/E-Token}\} \\
-\infty, & \text{if } j > i
\end{cases}
\end{aligned}
\end{equation}
\vspace{-5pt}

In addition, considering that different types of tokens possess distinct characteristics, the HSD module assigns independent normalization layers and feed-forward networks (FFNs) to each token type, projecting them into their own semantic subspaces to fully capture the semantic diversity of heterogeneous sequences.
Meanwhile, we introduce the Mixture-of-Recursions (MoR) mechanism ~\cite{bae2025mixture}, which increases the model’s depth and reasoning capacity without adding extra parameters.
To further enhance reasoning ability, the model incorporates external knowledge from a fine-tuned Large Language Model (LLM), where the LLM generates a textual “thought process” about users’ potential interests, which is then tokenized and integrated into the intent embeddings to strengthen semantic understanding and reasoning capability.

\paragraph{\textbf{Progressive Token-wise Decoder (PTD)}}
Given intent embeddings generated by the HSD, the PTD Module acts as the secondary decoder and adapts a traditional Transformer decoder architecture to generate the target item. Although intent embeddings encapsulate comprehensive representations of user behaviors, they may contain redundant information, which can adversely affect the accuracy of item prediction.
To address this, the PTD Module adopts a novel \textit{"Thinking-Refining-Generation"} paradigm for predicting the next item’s semantic ID. Specifically, the PTD Module utilizes a cross-attention mechanism in which the intent embeddings serve as both the keys and values. Based on these intent embeddings, we firstly compel the PTD Module to generate $K$ thinking tokens, which are designed to distill essential information and filter out irrelevant components from the intent embedding.

Moreover, inspired by recent research on reasoning in LLM~\cite{madaan2023self,wu2024large}, we extend PTD by integrating a \textit{refining module} to further strengthen its cognitive abilities and generative capacities. 
As shown in Fig. ~\ref{fig:gpr_v3_1} (c), the refining module is designed upon the diffusion paradigm~\cite{rombach2022high}, which consists of a noise generator and a reverse process modeled as a Markov chain. In the reverse process, the noise is iteratively removed by a conditional denoising module with a transformer architecture. Prefix thinking tokens are aggregated using \texttt{Sum\_Pooling} and serve as the condition for the denoising module. Consequently, the refining module refines the initial reasoning results, which are subsequently passed to downstream modules.

Finally, leveraging both the thinking tokens and the refining token, the PTD generates a sequence of semantic codes to represent the next item. In the inference, we further utilize the Trie-Constrained Value-Guided Beam Search to efficiently decode the accurate results. We will introduce details in Sec.~\ref{sec:vg-tb-bs}.

\paragraph{\textbf{Hierarchical Token-wise Evaluator (HTE)}}

Unlike traditional content recommendation systems, online advertising systems must jointly optimize user engagement and platform revenue.
This requires predicting multiple business metrics for each candidate advertisement, including click-through rate (CTR), conversion rate (CVR), and effective cost per mille (eCPM). 
To enable end-to-end optimization, these multi-faceted predictions must be aggregated into a single scalar objective that balances user experience and business goals. 
We refer to this aggregated metric as \texttt{final\_value}, which serves as the main optimization target throughout our system. 
The detailed formulation of \texttt{final\_value} and its components are provided in Eq.~\ref{eq:final_reward} in the reinforcement learning section.
To facilitate a truly end-to-end ad generation solution, we build an integrated value estimation module (HTE) upon the hierarchical GPR model that combines generation with value estimation. This module facilitates the generation of candidate ads, followed by the estimation of \texttt{final\_value}. 
This integrated, end-to-end approach offers significant advantages over traditional multi-stage systems: it enhances the consistency at the representation and objective levels, thereby mitigating conflicts between retrieval and ranking stages.
Additionally, this approach improves the overall computational efficiency of the advertising system, as shown below.
Beyond its role in value prediction during inference, HTE subsequently serves as the critic model in reinforcement learning post-training, enabling value-based advantage estimation for policy optimization.

\subsection{Value-Guided Trie-Based Beam Search}
\label{sec:vg-tb-bs}

The predicted semantic codes for target item by PTD may be invalid or suboptimal in advertising scenarios, such as those that map to items that do not actually exist, items whose geo-targeting constraints exclude the current user, items whose budget has run out, or items with low value. Although traditional beam search with post-filtering and post-ranking can remove the invalid results and rank the others by value, it incurs prohibitive computational costs and latency. Therefore, we propose \textit{Value-Guided Trie-based Beam Search}, which integrates trie constraints generated by user profiles and value estimations directly into the decoding step to evaluate the prefixes early. Specifically, given predicted values for each semantic code by HTE, we dynamically adjusts the beam width, with larger values corresponding to a wider beam for next semantic code, to improve the potential revenue. Then, we prune the search space via a Trie Tree generated by current user profiles. This Trie Tree is constructed by applying user targeting strategies in advertisement system to filter candidates according to attributes such as age and gender, and therefore only contains candidates that are consistent with the user's attributes, thereby enabling early user-level targeted filtering.

%% file: 4_Training_GPR_section_v2.tex
\begin{figure*}
    \centering
    \includegraphics[width=1.0\linewidth]{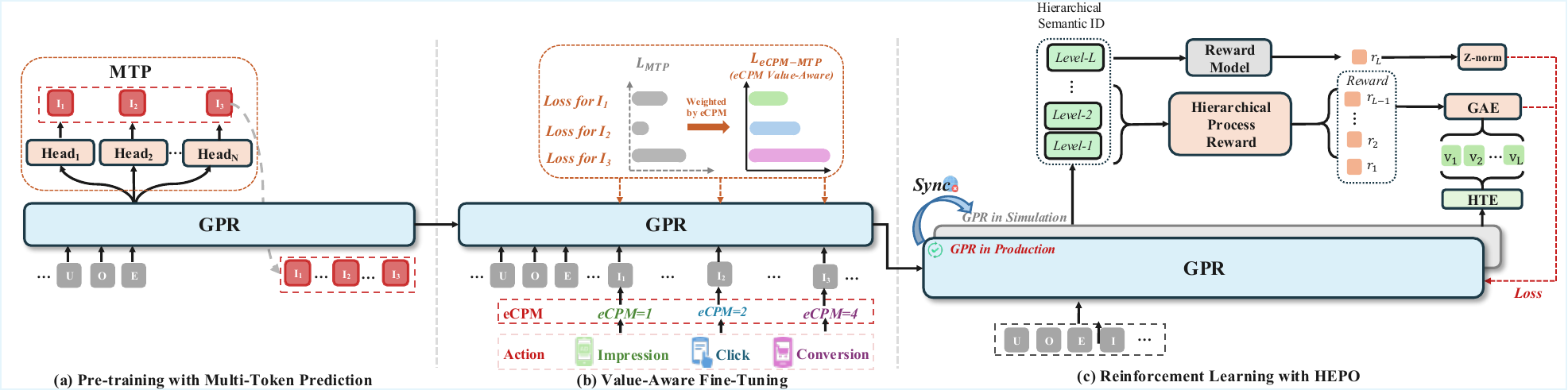}
    \caption{Training Pipeline of GPR.}
    \label{fig:training}
\end{figure*}

\section{Multi-Stage Training}

The GPR model is trained with a three-stage regimen designed for advertising recommendation under sparse signals, multi-business objectives, and a dynamic item space. 
First, pretraining uses multi-token prediction (MTP) to capture global, multi-interest user patterns. 
Next, an alignment stage introduces a value-aware MTP loss, reweighting updates toward higher-value items to align optimization with business priorities. 
Finally, reinforcement learning with HEPO in simulation enables exploration beyond logged exposures and surfaces under-served, high-value candidates under distribution shift. An overview of the full pipeline is shown in Fig. ~\ref{fig:training}.

\subsection{Pre-training with Multi-Token Prediction}

The pretraining stage injects advertising–scenario knowledge into GPR under sparse interaction signals (unlike dense organic traffic) and targets global, multi-interest user modeling. 
Users often pursue several intent threads in parallel; the objective here is to represent these threads jointly and coherently.
A large-scale industrial corpus from Tencent’s advertising platform is used, spanning one year and hundreds of millions of anonymized users and comprising both ad interactions (impressions, clicks, conversions) and organic engagements. 
For each user, a chronological sequence is constructed with the unified four-token schema (U/O/E/I-Token as in Sec.~\ref{sec:input}).

Items (ads) are encoded as $L$ coarse-to-fine semantic codes obtained via residual vector quantization, which provides a hierarchical, compact representation of item semantics. 
As a baseline, next-token prediction (NTP) conditions on the user history and context to predict the subsequent code. 
While effective for single-path dynamics, NTP implicitly assumes a dominant interest trajectory and tends to average over parallel interests, limiting coverage in advertising scenarios. To capture multiple concurrent interests, as shown in Fig. ~\ref{fig:training} (a), we employ MTP \cite{gloeckle2024better} by extending decoder with \(N\) parallel heads (default \(N=4\)). Each head independently predicts a complete \(L\)-level code path for one interest dimension, using the same backbone states but separate projection layers. This design enables concurrent modeling without mutual interference and preserves level-wise legality via masked decoding on each head.
 
The pretraining objective aggregates per-head, per-level likelihoods with simplex-constrained head weights \(\omega_j\) (\(\sum_j \omega_j=1\)), adaptively tuned to prioritize high-quality interest threads:
\begin{equation}
L_{\text{MTP}} = -\sum_{j=1}^{N} \sum_{t=1}^{T} \sum_{\ell=1}^{L} \omega_j^H \cdot \log P_j\left(I_{j,t,\ell} \mid S, C, I_{j,t,1:\ell-1}\right),
\label{eq:mtp_loss}
\end{equation}  

where $I_{j,t,\ell}$ denotes the \(\ell\)-th semantic code emitted by head \(j\) at position \(t\); \(P_j(\cdot)\) is the masked conditional probability on the legal set for that level; \(S\) and \(C\) denote sequence history and contextual features. We initialize \(\omega_j{=}1/N\) and allow subsequent stages to adjust the effective emphasis across heads via downstream supervision. This objective yields a backbone that encodes broad, disentangled interest structure and provides a stable foundation for subsequent business alignment and reinforcement learning.

\subsection{Value-Aware Fine-Tuning (VAFT)}

This stage bridges multi-interest pretraining and monetization goals by injecting action value and eCPM awareness into the MTP framework, so the model prioritizes high-value ads while preserving relevance. Although effective for multi-interest capture, vanilla MTP is misaligned with advertising objectives: (i) it assigns equal loss weight to ads with very different economic value, allowing low-eCPM long-tail items to dominate gradients; and (ii) it treats action types (impression, click, conversion) uniformly, ignoring their hierarchical business value (conversion $>$ click $>$ impression).

We introduce a per-head, per-position weight $\omega_{j,t}^{V}$ that encodes business value by combining the action type and the ad’s eCPM, as shown in Fig. ~\ref{fig:training} (b). The weight differentiates actions according to their value hierarchy (conversion $>$ click $>$ impression) and scales with a normalized eCPM to avoid magnitude distortion and spurious gradients. This term complements the head-level importance $\omega_j^{H}$ from pretraining, which prioritizes high-quality interest threads.
The value-aligned MTP loss multiplies the head importance and the action/eCPM weight:
\begin{equation}
L_{\text{eCPM-MTP}}
= -\sum_{j=1}^{N}\sum_{t=1}^{T}\sum_{\ell=1}^{L}
\big(\omega_j^{H}\,\omega_{j,t}^{V}\big)\,
\log P_j\!\big(I_{j,t,\ell}\mid S, C, I_{j,t,1:\ell-1}\big),
\label{eq:ecpm-mtp loss}
\end{equation}  
where the composite weight $\omega_j^{H}\omega_{j,t}^{V}$ integrates head-level interest quality with position-level business value, biasing updates toward high-eCPM actions while preserving multi-interest coverage. 
Other symbols follow the same definition in Eq.~\ref{eq:mtp_loss}.
In practice, $\omega_{j,t}^{V}$ uses calibrated or normalized eCPM with monotonic transforms, optional clipping to mitigate outliers, and fixed action-type coefficients to enforce the value hierarchy. This yields stable gradients, avoids overweighting rare spikes, and improves alignment with revenue objectives. 
Denominators are set to align $\omega_{j,t}^V$ with advertising business value (conversion > click > impression), as follows:
 
\textbf{Impression ($i=1$):} Denominator = 1 → $\omega_{j,t}^V \propto \text{eCPM}$ (basic revenue contribution); 

\textbf{Click ($i=2$):} Denominator = $\text{pCTR}$ → $\omega_{j,t}^V \propto \frac{\text{eCPM}}{\text{pCTR}}$ (reward ads with high click quality);

\textbf{Conversion ($i=3$):} Denominator = $\text{pCTR} \times \text{pCVR}$ → $\omega_{j,t}^V \propto \frac{\text{eCPM}}{\text{pCTR} \times \text{pCVR}}$ (prioritize ads driving actual conversions).


\subsection{Post-training with HEPO}

Supervised pretraining establishes a strong foundation by learning from historical exposure and engagement logs, but likelihood-only training faces an inherent limitation: logged data provides limited action coverage. The model observes only sequences generated by the historical policy, leaving many plausible high-value alternatives unexplored. Without counterfactual evaluation, the policy remains constrained to imitate past decisions rather than discovering superior strategies. Reinforcement learning addresses this limitation by constructing a high-fidelity simulation environment that enables counterfactual evaluation of policy-generated sequences, expanding action support beyond the historical distribution and allowing the model to explore new candidate ads before deployment.

In the setting of generative recommendation, the state $s$ includes the user’s interaction history, contextual signals such as device, time, and scene, the multi-level codes already emitted, and the level-specific legal masks. 
The action $a$ is a hierarchical decision produced by a $L$-level decoder (coarse→fine): at each level, the policy selects one quantized code from that level’s legal candidate set, and the final level resolves to the concrete ad to expose. 
The reward is assigned to the final decoding step that determines exposure, with optional light shaping (small fractions of the same signal) applied to earlier levels to preserve coarse-level intent. 
The episode corresponds to a single request or session and in the single-exposure case ends immediately after the final-level decision.

The GPR model consists of a Heterogeneous Sequence-wise Decoder (HSD) that produces intent embeddings $h=\mathrm{HSD}_\theta(s)$ from user context $s$.
The Progressive Token-wise Decoder (PTD) performs hierarchical decoding to generate action probabilities $\pi_\theta(z_\ell)$ over semantic tokens $z_\ell$ at each level $\ell$ where $\theta$ comprises the parameters of both HSD and PTD.
The Hierarchical Tower for eCPM estimation (HTE) serves as the value function $V_\phi$ in RL training, computing expected returns $v_\ell = V_\phi(s, z_{1:\ell-1})$, where $z_{1:\ell-1} = \{z_1, \ldots, z_{\ell-1}\}$ denotes the tokens selected at earlier levels.
The value function operates on $\mathrm{stopgrad}(h)$, ensuring that value gradients do not update the backbone and maintaining training stability.

\paragraph{\textbf{Reward Generation with simulation environment}}
Reinforcement learning requires reward signals for every candidate sequence the policy generates, but evaluating all candidates with the production ranking model in real-time is computationally prohibitive and would introduce unacceptable serving latency. 
Instead, we construct a high-fidelity simulation environment that replicates the production serving system for offline reward evaluation. 
The simulation environment is built upon production snapshots captured with small lag, preserving the complete infrastructure including retrieval indices, feature processing pipelines, and business constraint rules. 
It integrates two model types with different update frequencies: the pCTR/pCVR ranking models are directly copied from production to ensure reward fidelity, while the GPR policy models are periodically pushed from the training pipeline to evaluate evolving policies.

For each user request context $s$, the simulator performs beam search with the deployed GPR model to generate $K$ candidate advertisements (typically $K=40$), where each candidate is obtained through hierarchical decoding across $L$ levels of semantic IDs.
Each candidate is evaluated by the ranking models to obtain its predicted reward:
\begin{equation}
\begin{aligned}
R &= \texttt{final\_value}(s, \{z_\ell\}_{\ell=1}^L) \\
&=\mathrm{eCPM}(s, \{z_\ell\}_{\ell=1}^L) + \,\frac{1}{N}\sum_{i=1}^{N}\alpha_i\,\mathrm{target}_i(s, \{z_\ell\}_{\ell=1}^L),
\end{aligned}
\label{eq:final_reward}
\end{equation}
where $\{z_\ell\}_{\ell=1}^L$ denotes the token sequence that uniquely identifies this candidate, $\mathrm{eCPM}$ is the predicted effective cost per mille, $\mathrm{target}_i$ are auxiliary objectives such as pCTR and pCVR with weights $\alpha_i$. 
The simulation environment records the generation probabilities $\pi_{\theta_{\text{old}}}(z_\ell)$ at each decoding level, which serve as behavior policy probabilities for off-policy correction.

\paragraph{\textbf{Hierarchical Process Rewards}}
When rewards are assigned only at the final exposure level, intermediate hierarchical decisions receive no direct feedback signal. 
This creates a credit assignment problem: coarse-level choices must rely solely on bootstrapped value estimates to assess their contribution to the terminal outcome, resulting in weak and high-variance learning signals at upper levels. 
Consider a user interested in smartphones who rejects a recommendation for "Mobile Phones → Brand A → Item X." 
Without intermediate rewards, the negative terminal signal affects all levels uniformly, incorrectly reducing the probability of the "Mobile Phones" category despite the user's genuine interest—the error lies at the brand level, not the category level. 
To address this, we propose HEPO (\textit{H}ierarchical \textit{E}nhanced \textit{P}olicy \textit{O}ptimization), which constructs process rewards that leverage user-specific preference patterns to provide direct supervision signals at each hierarchical level.

For each level $\ell$, we derive a per-token popularity score $P_\ell(t) \in [0,1]$ from the user's successful historical interactions, representing how frequently token $t$ appears in recommendations that led to positive outcomes. 
For the chosen token $z_{\ell}$ at level $\ell$, we compute a preference signal by comparing its popularity against the average popularity of all legal candidates $\mathcal{S}_{\ell}$:
\begin{equation}
\Delta_{\ell} = P_\ell(z_{\ell}) - \frac{1}{|\mathcal{S}_{\ell}|} \sum_{t \in \mathcal{S}_{\ell}} P_\ell(t).
\end{equation}
The baseline subtraction ensures zero-mean signals when all candidates have equal popularity, preventing systematic bias. The step reward at each level is then:

\begin{equation}
r_{\ell} = 
\begin{cases}
\alpha_\ell \max(0, \Delta_{\ell}), & \ell < L,\\
R, & \ell = L,
\end{cases}
\label{eq:process reward}
\end{equation}
where $R$ is the reward obtained from the simulator as defined in Eq.~\ref{eq:final_reward}, $\alpha_\ell$ are small scaling factors that ensure process rewards guide learning without overwhelming the terminal reward.

\paragraph{\textbf{Advantage and loss}}
At intermediate levels, process rewards must be connected to long-term business outcomes through a learned value function that enables proper credit assignment across the hierarchy. 
At the final exposure level, terminal rewards exhibit high variance across requests, making critic-based value estimation unstable. 
Since the exposure decision directly determines the displayed advertisement and requires stable optimization, we instead use within-request z-score normalization. 
For each request, the simulation environment generates $K$ candidates and their relative advantage is assessed by normalizing rewards within this set.

For coarse levels $\ell < L$, we compute advantages via Generalized Advantage Estimation (GAE) using the process rewards from Eq.~\ref{eq:process reward}. 
The cumulative return $G_\ell = \sum_{k=0}^{L-\ell} \gamma^k r_{\ell+k}$ and temporal difference errors $\delta_\ell = r_\ell + \gamma  V_\phi(s, z_{1:\ell}) -  V_\phi(s, z_{1:\ell-1})$ are computed, where $V_\phi(\cdot)$ is the value function and  $\gamma$ is the discount factor.
For the final level, z-score normalization is applied over the over the $K$ candidates. 
The advantage at each level is:
\begin{equation}
A_{\ell} = 
\begin{cases}
\sum_{l=0}^{L-\ell-1} (\gamma\lambda)^l \delta_{\ell+l}, & \ell < L,\\[8pt]
\frac{R - \mu_K}{\sigma_K + \epsilon}, & \ell = L,
\end{cases}
\end{equation}
where $\lambda$ controls the bias-variance tradeoff for GAE, $R$ is the terminal reward from Eq.~\ref{eq:final_reward}, and $\mu_K$ and $\sigma_K$ are the mean and standard deviation computed over candidates.

The policy model with parameters $\theta$ is updated by minimizing:
\begin{equation}
\begin{aligned}
\mathcal{L}_{\theta} = \mathbb{E}\Bigg[&\sum_{\ell=1}^{L} c_\ell \min\Big(\rho_{\ell} A_{\ell}, \text{clip}(\rho_{\ell}, 1-\epsilon, 1+\epsilon) A_{\ell}\Big)\Bigg], \\
\end{aligned}
\end{equation}
where $\rho_{\ell} = \pi_\theta(z_\ell) / \pi_{\theta_{\text{old}}}(z_\ell)$ is the importance ratio and $\pi_{\theta_{old}}$ is the behavior policy used for sampling in simulation. 
The value function is trained with mean squared error across all levels:
\begin{equation}
\mathcal{L}_{\phi} = \mathbb{E}\left[\sum_{\ell=1}^{L}( V_\phi(s, z_{1:\ell-1}) - G_\ell)^2\right].
\end{equation}

\paragraph{\textbf{Anticipatory Request Rehearsal (ARR)}}
The advertising ecosystem is highly dynamic: user interests evolve and premium creative inventory changes daily. A model trained solely on historical data risks being reactive rather than proactive. 
To enable anticipatory adaptation, we introduce Anticipatory Request Rehearsal (ARR), which generates synthetic training samples that approximate users' future request states.

Rather than using stale historical contexts, ARR constructs synthetic requests based on each user's current state to better approximate their next actual interaction. 
The sampling frequency adapts to user activity patterns: for high-activity users with frequent requests, synthetic samples are generated every 2 hours during peak periods  and every 4 hours during off-peak periods; for low-activity users, the intervals are adjusted proportionally to their request rates. 
Each synthetic request is constructed from the user's current state as follows. 
The Organic token is reconstructed using the user's most recently viewed organic content, including short videos and articles, to reflect evolving interests.
The User token is directly reused from the previous request for high-activity users, as their profile features remain stable over short horizons. 
The Environment token, which includes predicted ad position, placement type, and privacy settings, is queried in real-time to capture the latest system state. 
These synthetic samples are processed identically to observed samples in the simulation environment: the deployed GPR model generates candidates, the ranking model evaluates them, and advantages are computed following standard procedures.

%% file: 5_Experiments_section.tex
\section{Experiments}

We validate GPR through a four-step progression. 
First, we assess the proposed RQ-KMeans+ tokenizer on real and organic ads, showing it learns compact, semantically coherent codes that support a unified representation space. 
Second, we study if the proposed HHD architecture outperform strong baselines (HSTU, OneRec) for modeling long-horizon, long-sequence heterogeneous user behaviors. 
Third, we examine whether the value-aligned training strategies (eCPM-aware MTP and the proposed HEPO) outperform NTP and DPO baselines on monetization-oriented objectives.
Finally, we benchmark the end-to-end GPR system in large-scale online A/B tests against a mature cascading pipeline, confirming practical economic gains and robust generalization in production.

\noindent
\noindent

\subsection{Multimodal Tokenization Performance}

\textit{Experiment setting.}
A large-scale corpus from Tencent’s advertising platform is curated, covering ads and organic media (short videos, social feeds, news). 
The corpus comprises heterogeneous multimodal signals—textual metadata (titles, tags, descriptions) and visual content (thumbnails and sampled frames) —aligned at the item and session level.
Near-duplicate samples are filtered to reduce redundancy and 
category distributions are balanced to limit sampling bias.
The corpus is partitioned into 80\% training and 20\% test. 
RQ-KMeans+ is compared against two established tokenizers: RQ-VAE~\cite{rq-vae} (standard discrete tokenizer prone to dead codes) and RQ-KMeans~\cite{rq-kmeans} (K-means initialization without VAE-style refinement). 
We introduce three metrics to assess code quality. Specifically, Collision Rate measures the fraction of items mapped to an identical code and lower values are better. 
Code Usage Rate at level~1 (CUR\(_{\mathrm{L1}}\)) measures the proportion of active codes and higher values are better. Path Average Similarity (PAS) measures the mean embedding similarity among items that share a code and higher PAS indicate more coherent collisions. 

\textit{Results and Analysis.}
Tab. ~\ref{tab:tokenizer} shows that RQ-KMeans+ achieves the best overall code quality. It yields the lowest collision rate (20.60\%), a relative reduction of 11.2\% vs. RQ-VAE (23.21\%) and 3.7\% vs. RQ-Kmeans (21.40\%). Code usage at level-1 remains near saturation (99.36\%, close to the 100\% of RQ-Kmeans and +7.2 points over RQ-VAE), indicating that the reduction in collisions is not obtained by leaving codes idle. 
Importantly, PAS reaches 0.992 (vs. 0.985/0.986), implying that the remaining collisions group items with higher semantic coherence. 
These results indicate that items grouped under the same semantic ID by RQ-Kmeans+ are more likely to belong to the same category and exhibit higher similarity, thus leading to more reasonable code collisions.

\begin{table}[tb!]
  \centering
  \caption{Performance of Different Tokenizers}
  \vspace{-8pt}
  \label{tab:tokenizer}
  \begin{adjustbox}{max width=\linewidth}
  \begin{tabular}{l c S[table-format=2.2] c S[table-format=1.3]}
    \toprule
    \textbf{Model} & \textbf{Collision (\%)$\downarrow$} & \textbf{CUR$_{L1}$ (\%)$\uparrow$} & \textbf{PAS$\uparrow$ } \\
    \midrule
    RQ-VAE                 & 23.21 & 92.13  & 0.985 \\
    RQ-Kmeans              & 21.40 & \textbf{100} & 0.986 \\
    \textbf{RQ-KMeans+ (Ours)}  & \textbf{20.60} & 99.36  & \textbf{0.992} \\
    \bottomrule
  \end{tabular}
  \end{adjustbox}
\end{table}

\begin{table}[tb!]
  \centering
   \caption{HHD Architecture Study}
  \label{tab:gpr-ablation-baseline}
  \vspace{-8pt}
  \begin{tabular}{@{}p{0.5\linewidth} S[table-format=1.0] c@{}}
    \toprule
    \textbf{Model} & {\textbf{HitR@100} (\%)} & \textbf{$\Delta$ vs. HSTU} \\
    \midrule
    \textbf{Baselines} \\
    \quad HSTU (Decoder-only) & 18.98 & -- \\
    \quad OneRec (Encoder-Decoder) & 19.85 & +4.6\% \\
    \midrule
    \textbf{HSD} \\
    \quad + Hybrid Attention & 20.56 & +8.3\% \\
    \quad + Token-Aware FFN & 21.98 & +15.8\% \\
    \quad + Token-Aware Layer Norm & 20.76 & +9.4\% \\
    \quad + Mixture of Recursions & 20.09 & +5.9\% \\
    \quad + External Knowledge & 20.13 & +6.1\% \\
    \midrule
    \textbf{PTD} \\
    \quad + Thinking & 21.75 & +14.6\% \\
    \quad + Refining & 19.61 & +3.3\% \\
    \midrule
    \textbf{HTE} \\
    \quad + HTE & 19.91 & +4.9\% \\
    \midrule
    \textbf{Training} \\
    \quad + Multi-Token Prediction & 22.38 & +17.9\% \\
    \midrule
    \textbf{GPR (Full)} \\
    \quad + All (HSD+PTD+HTE) & 27.32 & +43.9\% \\
    \bottomrule
  \end{tabular}
  \vspace{-10pt}
\end{table}

\subsection{User Behavior Modeling Performance}
\label{sec:User Behavior Modeling Performance}

\textit{Experiment setting.} We use a large-scale Tencent advertising dataset  
(user interactions with ads and organic content).
Training uses one year of anonymized user interactions and validation uses the next calendar day.
Histories are serialized in strict time order with the unified four-token schema (U/O/E/I).
The same tokenizer (RQ-KMeans+) and feature pipelines are shared by all models.
Baselines are HSTU~\cite{zhai2024actions} (decoder-only) and OneRec~\cite{zhou2025onerec2} (encoder–decoder).
The task is to predict the next interacted item, and each model is required to emit top 100 results from the full catalog snapshot, which contains millions of items.
HitRate@100 is defined by this retrieval protocol: a hit is recorded if the ground-truth ad appears among the top-100 candidates.
This metric avoids overly strict code equality and evaluates whether the generated semantic code localizes the correct item in embedding space.

\textit{Results and Analysis.}
Tab. ~\ref{tab:gpr-ablation-baseline} reports the headline results. Full GPR with HHD attains 27.32\% HitR@100, a relative gain of +43.9\% over HSTU and +37.6\% over OneRec. OneRec itself exceeds HSTU because it ingests non-purely sequential inputs and aggregates richer feature views, which supplies stronger conditioning than a strictly autoregressive decoder. The trade-off is weaker universality at inference time, since dependence on pre-structured fields and candidate schemas reduces flexibility compared with a decoder that can condition on any serialized history.

Ablations attribute the improvements to three components.
HSD strengthens user–intent encoding on long, heterogeneous histories. Hybrid Attention relaxes the causal mask on the non-generative prefix and adds +8.3\%. Token-Aware FFN/LN reduces cross-type interference and yields relative gains of +15.8\% and +9.4\%. 
Mixture of Recursions increases effective depth with shared parameters and contributes +5.9\%. 
Injecting “thought-process” tokens from the LLM further enriches the intent representation for a +6.1\% gain. 
PTD improves generation quality and diversity. Implicit thinking enforces multi-step latent refinement and delivers +14.6\%. Refining can bring 3.3\% improvement. HTE 
improves HitR@100 to 19.91\%, showing that explicit value prediction sharpens candidate ordering and better prepares the generative system for downstream auction. Multi-Token Prediction captures parallel interests within a single pass and provides the largest gain at +17.9\%, supporting the multi-threaded-interest hypothesis.

\begin{figure}[t!]
    \centering
    \includegraphics[width=1.0\linewidth]{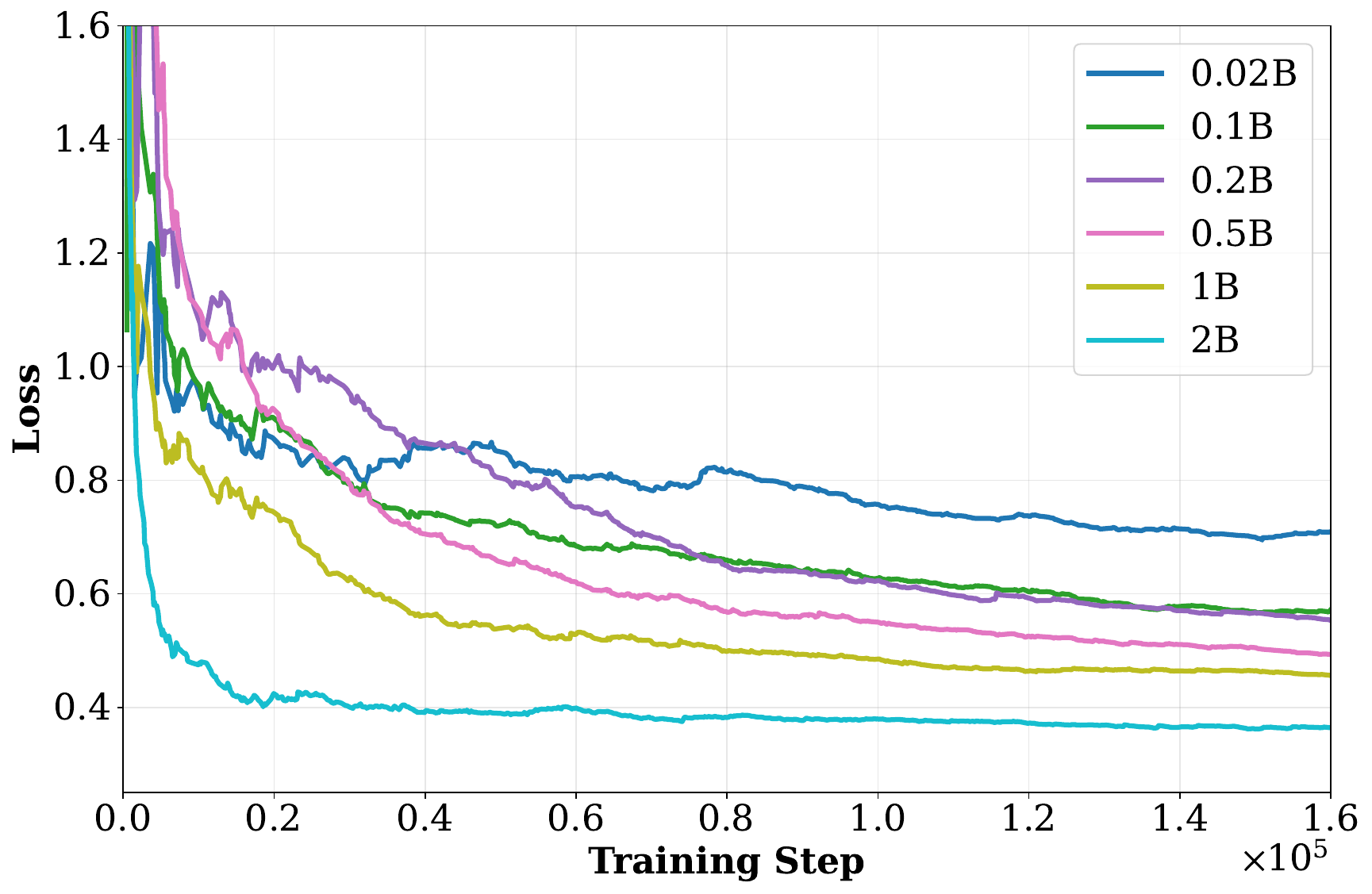}
    \vspace{-15pt}
    \caption{Comparison of loss curves for six different GPR parameter sizes.}
    \label{fig:scaling}
\end{figure}

\textit{Scaling.}
We investigate the scaling properties of GPR across six dense parameter sizes: 0.02B, 0.1B, 0.2B, 0.5B, 1B, and 2B. Notably, unlike typical large language models, GPR's total size is dominated by sparse parameters, which total approximately 80B. Fig. ~\ref{fig:scaling} presents the evolution of the training loss for these six model variants. The results conclusively demonstrate a robust scaling law: models with greater parameter counts consistently achieve lower loss values as training progresses. This empirical observation validates the substantial potential for performance enhancement gained by scaling up the model size.

\begin{table}[tb]
  \centering
  \caption{Training \& Alignment Study}
  \vspace{-8pt}
  \label{tab:objective-alignment}
  \begin{tabular*}{1\linewidth}{@{\extracolsep{\fill}} l c c c c @{}}
    \toprule
    \multirow{2}{*}{\textbf{Model}} & \multirow{2}{*}{\textbf{nDCG}} & \multirow{2}{*}{\textbf{OPR}} & \textbf{Avg} & \textbf{Max} \\
    & & & \textbf{\texttt{final\_value}} & \textbf{\texttt{final\_value}} \\
    \midrule
    \multicolumn{5}{l}{\textbf{Pretraining \& Fine-tuning}} \\
    MTP (base) & 0.3868 & 0.5292 & 0.2412 & 0.6201 \\
    + VAFT    & 0.3925 & 0.5348 &  -- & -- \\
    \midrule
    \multicolumn{5}{l}{\textbf{Post-training}} \\
    + DPO          & 0.4383 & 0.5463 & 0.2442 & 0.6659 \\
    + HEPO         & 0.4413 & 0.5509 & 0.2630 & 0.7619 \\
    \bottomrule
  \end{tabular*}
  \vspace{-10pt}
\end{table}

\subsection{Business Alignment Performance}

\textit{Experiment setting.}
This study evaluates whether the joint training strategy—MTP pretraining, eCPM-aware fine-tuning, and HEPO—
improves monetization alignment. 
The dataset follows Section~\ref{sec:User Behavior Modeling Performance}: one year for training and the subsequent day for validation. In addition to sequence signals, labels include eCPM and action types (impression, click, conversion) to reflect value hierarchy. 
Methods compared are: MTP, a baseline multi-token prediction model trained with uniform head/position weights under likelihood only;  MTP+VAFT, a business-aligned variant that reweights the MTP loss at each position using a value-aware factor; MTP+DPO~\cite{rafailov2023direct}, a preference-based fine-tuning that learns from pairwise orderings constructed under predicted rewards; and MTP+HEPO. We evaluate with nDCG (normalized Discounted Cumulative Gain), which measures ranked-list quality with higher
weight on the top, and OPR (Ordered Pair Ratio), which measures
the fraction of item pairs ordered correctly. Both metrics are computed under the eCPM-oriented ranking.

We also report the average and maximum \texttt{final\_value} computed from the simulation environment as defined in~\ref{eq:final_reward} on a fixed calendar day.
For each request, the simulator generates $K$ candidates.
We collect all \texttt{final\_value} scores across all candidates and all requests, and apply min-max normalization to this complete set, as \texttt{final\_value} can be both positive and negative.
For confidentiality, we report only the normalized values rather than the original scale.
We then compute the mean \texttt{final\_value} (averaged over the $K$ candidates) and max \texttt{final\_value} (maximum over the $K$ candidates) for each request using the normalized values, and aggregate these statistics across all evaluation requests.
Note that simulated \texttt{final\_value} improvements do not directly translate to online revenue gains.

\textit{Results and analysis.}
The results of business alignment performance are shown in Tab. ~\ref{tab:objective-alignment}.
Relative to MTP, MTP+VAFT raises nDCG from 0.3868 to 0.3925 and OPR from 0.5292 to 0.5348 by reweighting the loss with action type and normalized eCPM, which shifts learning toward high-value impressions while retaining relevance. 
Reinforcement learning evaluates policy-generated sequences in simulation, aligning supervision with sequence-level rewards and enabling safe counterfactual exploration. 
DPO optimizes pairwise preferences that favor higher-value items under matched context, sharpening local orderings and improving nDCG to 0.4383, with normalized average \texttt{final\_value} increasing from 0.2412 to 0.2442 and max \texttt{final\_value} from 0.6201 to 0.6659. 
HEPO further surpasses DPO, achieving nDCG 0.4413, OPR 0.5509, average \texttt{final\_value} 0.2630, and max \texttt{final\_value} 0.7619.

\subsection{Online Performance}

\begin{table}[tb!]
  \centering
  \caption{Online A/B Test Results }
  \vspace{-8pt}
  \label{tab:ab-online}
  \small
  \begin{tabular}{p{3.0cm}ccc}
    \toprule
    \textbf{Version} & \textbf{GMV} & \textbf{GMV-Normal}  & \textbf{Costs} \\
    \midrule
    \multicolumn{4}{l}{\textit{Launches with incremental changes.}} \\
      v0.1: HSD+NTP+DPO  & +2.11\% & +2.42\%  & +3.29\% \\
      v0.2: +HEPO w/o ARR       & +0.70\% & +0.67\%  & +0.36\% \\
      v0.3: +MTP+Thinking  & +0.63\% & +0.94\%  & +0.21\% \\
      v0.4: +PTD           & +0.71\% & +1.04\%  & +0.12\% \\
      v0.5: +HEPO w/ ARR & +0.58\% & +0.81\%  & +0.23\%\\ 
    \bottomrule
  \end{tabular}
  \vspace{-3pt}
\end{table}

\begin{table}[tb!]
\centering
\caption{Stratified Analysis of Online A/B Test Results }
\vspace{-8pt}
  \label{tab:ab-online-seg}
\small
\begin{tabular}{llllll}
\toprule
  &   & \textbf{GMV} & \textbf{CTR} & \textbf{CVR} & \textbf{CTCVR} \\
\midrule
v0.1 &  & +2.11\% & +1.69\% & +1.15\% & +3.16\% \\
\midrule
\multirow{5}{*}{User Group} & UG1 & +3.56\% & +2.51\% & +0.82\% & +3.72\% \\
& UG2 & +3.84\% & +2.06\% & +1.30\% & +3.80\% \\
& UG3 & +0.92\% & +2.18\% & +1.91\% & +4.63\% \\
& UG4 & +0.45\% & +1.08\% & +1.53\% & +2.87\% \\
& UG5 & +3.68\% & +0.05\% & +0.32\% & +0.50\% \\
\hline
\multirow{2}{*}{Ad Group} & new & +2.97\% & +2.25\% & +1.41\% & +4.02\% \\
& non-new & +1.65\% & +1.42\% & +1.12\% & +2.78\% \\
\bottomrule
\end{tabular}
\vspace{-10pt}
\end{table}

\textit{Experiment setting.}
We evaluate and deploy GPR on Weixin Channels, a production advertising platform with hundreds of millions of active users and tens of millions of dynamic ads. 
The comparison baseline is a mature multi-stage cascade with multiple retrieval methods and customized strategies. 
Primary online metrics include GMV, Costs, and CTCVR, with GMV as the primary KPI  because it directly reflects business return. We conduct five sequential online A/B evaluations of the GPR stack. Each iteration undergoes rigorous validation and produces stable, statistically significant incremental improvements over the preceding version on the primary KPIs.

\begin{itemize}
  \item \textbf{v0.1:} \textbf{HSD} + NTP + DPO (initial GPR deployment).
  \item \textbf{v0.2:} \textbf{HEPO} w/o \textbf{ARR} (Anticipatory Request Rehearsal).
  \item \textbf{v0.3:} Add \textbf{MTP} (multi-token prediction) and \textbf{Thinking}.
  \item \textbf{v0.4:} Introduce \textbf{PTD} (progressive token-wise decoder).
  \item \textbf{v0.5:} \textbf{HEPO} w/ \textbf{ARR} (Anticipatory Request Rehearsal).
\end{itemize}

\textit{Results and Analysis.}
Tab. ~\ref{tab:ab-online} summarizes the online effects across five evaluations.
The first full-scale launch (HSD+NTP+DPO) establishes the baseline lift with +2.11\% GMV and +3.29\% Costs.
The second major rollout replaces DPO with HEPO (without anticipatory request rehearsal), delivering an additional +0.70\% GMV and +0.36\% Costs.
Subsequent large-scale experiments confirm steady, incremental gains: introducing MTP+Thinking contributes +0.63\% GMV, adding PTD contributes +0.58\% GMV, and applying HEPO with an anticipatory request rehearsal contributes +0.58\% GMV.
Across all rounds, both GMV and GMV-Normal (ads optimized for clicks or conversions, which accounts for the majority of total GMV) increase consistently, indicating stronger monetization under unchanged latency and stability constraints.

The stratified analysis based on first launch (v0.1) in Tab. ~\ref{tab:ab-online-seg} shows consistent gains across users and ad segments. 
UG1 and UG2 are low-activity users and show strong gains, for example UG1 reaches GMV +3.56\%, CTR +2.51\%, CVR +0.82\%, and CTCVR +3.72\%. 
Mid-activity groups UG3 and UG4 also improve across engagement and efficiency, with UG3 showing the largest CTCVR lift at +4.63\%. 
UG5 denotes high-activity users and exhibits smaller CTR and CVR changes yet still delivers GMV +3.68\%, indicating better allocation toward higher-value ads even for heavy users. 
For inventory, newly launched  ads ($\leq$3 days) outperform established ads (>3 days ) with GMV +2.97\% vs. +1.65\% and CTCVR +4.02\% vs. +2.78\%.
This pattern indicates stronger cold-start handling while preserving gains for mature inventory.

%% file: 2_Related_Work_section_v2.tex
\section{Related Work}

This section reviews prior work across four core areas relevant to GPR: LLMs as generative rankers, generative recommendation models, end-to-end recommendation frameworks, and the application of Reinforcement Learning (RL).

\textbf{LLMs as Generative Rankers.}
In the last two years, researchers have begun exploring the application of LLMs in recommender systems, primarily leveraging their ability to generate the next item recommendation~\cite{wu2024survey}. In these efforts, user behavior sequences are typically transformed into text sequences, and the LLMs are tasked with generating the next item based on this textual input~\cite{zhu2024collaborative, lin2024rella, li2024leadre, zheng2024adapting, harte2023leveraging, zhang2023chatgpt}. However, a core limitation of this approach is the fixed vocabulary of LLMs, which hinders their ability to adapt effectively to the dynamically changing and large-scale item sets characteristic of modern advertising recommendation scenarios.

\textbf{Generative Recommendation Models.}
Recently, driven by the success of LLMs, there has been a significant shift towards generative recommendation~\cite{rajput2023recommender,chen2024hllm,zhai2024actions,liu2024kuaiformer,badrinath2025pinrec,yang2024unifying,qiu2025one,huang2025towards,wang2025scaling,jiang2025large}. TIGER~\cite{rajput2023recommender} is a generative recommendation model that leverages semantic ids, using a sequence-to-sequence framework to generate such semantic ids for recommendations. HLLM \cite{chen2024hllm} leverages the pre-trained capabilities of LLMs and models item representations and user interests respectively via a two-layer architecture. HSTU~\cite{zhai2024actions} adopts a decoder-only architecture to process ultra-long user histories and generate item recommendations. These work have demonstrated the potential of generative models to unify different recommendation tasks and handle long user sequences. MTGR~\cite{han2025mtgr} adopts the HSTU architecture while retaining the original deep learning recommendation model (DLRM) features. COBRA~\cite{yang2025sparse} integrates sparse semantic IDs and dense vectors through a cascading process. 
GPR builds on this generative trend but introduces a novel HHD architecture. Unlike previous approaches which entangle user representation with item generation, GPR decouples these two tasks via a hierarchical structure and propose a novel "understanding-thinking-refining-generation" paradigm.

\textbf{End-to-End Recommendation Frameworks.}
Recent studies have attempted to construct more unified, end-to-end recommendation systems, such as OneRec\cite{zhou2025onerec}. 
OneRec unifies retrieve-and-rank via a encoder-decoder architecture and a preference alignment algorithm (e.g., DPO) in a video recommendation system. By contrast, GPR is the first end-to-end generative solution successfully deployed in large-scale advertising system, which faces unique challenges including behavior heterogeneity and sparsity, multi-objective optimization, precise value prediction, etc. 

\textbf{RL in Recommender Systems.}
In traditional recommender systems, reinforcement learning (RL) is used to optimize slate/page decisions and long-horizon value, rather than just next-item accuracy. Seq2Slate~\cite{bello2018seq2slate} formulates re-ranking as autoregressive slate generation and trains via policy gradients, enabling the model to assign credit to the entire slate (instead of treating items independently) and capture cross-position effects that pointwise losses miss. SlateQ~\cite{ie2019reinforcement} shows that, under mild user-choice assumptions, the long-term value of a slate can be decomposed into itemwise terms—making TD/Q-learning tractable for page-wise optimization and enabling production-scale experiments. At an extremely large scale, YouTube’s system introduces Top-K off-policy correction to learn the recommendation policy from logged implicit feedback, while incorporating a learned model of the behavior policy to adjust data biases. Ad-mixed feeds further highlight the need for RL, as the system must jointly balance platform revenue, advertiser ROI, and user experience at the slate level: DEAR~\cite{zhao2021dear} addresses ad insertion as a joint decision (whether to insert an ad, which ad to show, and where to place it) using a DQN that optimizes long-run value under the delayed/mixed rewards typical of advertising. To reduce online risk, teams often start with offline RL; a common choice is Conservative Q-Learning (CQL)~\cite{kumar2020conservative}, which penalizes out-of-distribution actions to prevent the learned value function from overestimating unseen choices before online fine-tuning.

\textbf{RL for Generative Recommendation.} In generative recommendation, RL is used to align sequence-producing policies with user and business preferences. GeMS~\cite{deffayet2023generative} learns a variational latent space for slates, enabling an RL agent (e.g., SAC) to act in this continuous space before decoding full lists—sidestepping combinatorial action spaces while still optimizing sequence-level rewards. When hand-crafted rewards are brittle, PrefRec~\cite{xue2023prefrec} first learns a reward model from human/trajectory preferences, then optimizes policies against this model to improve long-term engagement. Direct Preference Optimization (DPO)-style objectives~\cite{rafailov2023direct} adapt this idea to sequence models (e.g., Softmax-DPO; DPO4Rec), turning pairwise preferences into stable training losses that can replace or complement RL during post-training (e.g., for aligning generative policies with user preferences). Recent industrial systems combine these elements end-to-end: OneRec-V1~\cite{zhou2025onerec} unifies retrieve-and-rank with a generator and iterative preference alignment (via DPO), reporting a +1.6\% online watch-time lift; OneRec-V2~\cite{zhou2025onerec2} shifts to a lazy decoder-only stack and adds RL on real user feedback (with duration-aware reward shaping and adaptive ratio clipping) to optimize long-term engagement.

%% file: 6_Conclusion_section.tex
\section{Conclusion}
In this paper, we present GPR (Generative Pre-trained Recommender) — the first one-model framework that formulates advertising recommendation as an end-to-end generative task, replacing the conventional multi-stage cascaded pipeline with a unified generative paradigm.
Unlike traditional systems that suffer from objective inconsistency and error accumulation across stages, GPR achieves global optimization and long-term value alignment within a single holistic model. Our framework introduces systematic innovations in three key aspects: unified representation, architectural design, and training strategy. These advancements effectively address long-standing challenges in industrial recommender systems, including heterogeneous data modeling, training–inference efficiency trade-offs, and insufficient long-term reward optimization. 

Extensive large-scale experiments and real-world online A/B testing demonstrate the superiority of GPR, achieving a significant improvements in GMV and CTCVR. Its fully generative and unified design further propels advertising recommendation systems from stage-wise optimization toward end-to-end intelligent decision-making, enabling the system to understand user intent in a more unified and adaptive manner, optimize long-term value, and continuously drive the intelligent evolution of the digital economy ecosystem.